\documentclass[useAMS,usenatbib]{mn2e}

\usepackage{natbib}
\usepackage{graphicx}
\usepackage{multirow}
\usepackage{subfig}

\title[The NGC7771+NGC7770 Minor Merger]{The NGC7771+NGC7770 Minor
  Merger: Harassing the Little One?\thanks{Based on observations
     collected at the Centro Astron\'omico Hispano-Alem\'an
    (CAHA) at Calar Alto, operated jointly by the Max-Planck Institut
    f\"ur Astronomie and the Instituto de Astrof\'{\i}sica de
    Andaluc\'{\i}a (CSIC). The data presented here were obtained [in
    part] with ALFOSC, which is
provided by the Instituto de Astrof\'{\i}sica de Andaluc\'{\i}a (IAA)
under a joint agreement with the University of Copenhagen and NOTSA. 
}}
\author[Almudena Alonso-Herrero et al.]
{\parbox{\textwidth}{Almudena Alonso-Herrero,$^{1,2}$
    F. Fabi\'an Rosales-Ortega,$^{3,4}$  Sebasti\'an F. S\'anchez,$^{5,4}$ Robert C. Kennicutt,$^6$
    Miguel Pereira-Santaella,$^{7}$ \'Angeles I. D\'{\i}az$^3$}
\vspace{0.4cm} \\
$^{1}$Instituto de F\'{\i}sica de Cantabria, CSIC-UC, 39005 Santander,
 Spain. E-mail: aalonso@ifca.unican.es \\
$^{2}$Augusto Gonz\'alez Linares Senior Research Fellow\\
$^{3}$Departamento de F\'{\i}sica Te\'orica, Universidad Aut\'onoma de
Madrid, 28049 Madrid, Spain\\
$^{4}$Centro Astron\'omico Hispano Alem\'an Calar Alto, CSIC-MPG,
04004 Almer\'{\i}a, Spain\\
$^{5}$Instituto de Astrof\'{\i}sica de Andaluc\'{\i}a, CSIC,
18080 Granada, Spain\\
$^{6}$Institute of Astronomy, University of Cambridge, Cambridge CB3 0HA\\
$^{7}$Istituto di Astrofisica e Planetologia Spaziali, INAF-IAPS, 
00133 Rome, Italy
}

\begin{document}

\date{Accepted --- . Received --- ; in original form --- }


\maketitle

\label{firstpage}

\begin{abstract}
Numerical simulations of minor mergers, typically having
mass ratios greater than 3:1,  predict little enhancement in the global
star formation activity. However, these models
also predict that the satellite galaxy  is more susceptible to the effects
of the interaction than the primary.  
We use optical integral field spectroscopy and deep optical imaging to
study the NGC~7771+NGC~7770 interacting system ($\sim$ 10:1
stellar mass ratio) to test these predictions.  We find that the
satellite galaxy NGC~7770
is currently experiencing a galaxy-wide starburst with most of the
optical light being from young and post-starburst stellar populations
($<1\,$Gyr). This galaxy lies off  of the local
star-forming sequence for composite galaxies 
with an enhanced integrated specific star formation rate. 
We also detect in the outskirts of NGC~7770 H$\alpha$
emitting gas filaments. This gas appears to
have been stripped from one of the two  galaxies and is being excited
by shocks. All these results are consistent with a minor-merger induced
episode(s) of star formation in NGC~7770 after the first close
passage. Such effects are not observed on the primary galaxy 
NGC~7771. 
\end{abstract}


\section{Introduction}
Minor mergers are defined to have mass ratios greater than
approximately 3:1. They have been increasingly recognized as important
players in galaxy evolution and, in particular, in the formation and
assembly of bulges especially in lower mass systems \citep[see][and
references therein]{Hopkins2010}. Numerical simulations of minor
mergers indicate that they can trigger nuclear activity and
transform the morphologies of galaxies \citep{Mihos1994,
  Hernquist1995, Naab2003, Robertson2006, Qu2011}. The simulations of
\citet{Cox2008} showed in detail that during the 
interaction process of minor mergers the global star formation rate
(SFR) only increases moderately and the increase 
is a function of the mass ratio of
the galaxies.  Furthermore, the fractional SFR enhancement is much
higher in the satellite galaxy as it is more susceptible to the
tidal forces induced by the interaction.

\begin{figure}

\hspace{0.7cm}
\resizebox{0.8\hsize}{!}{\rotatebox[]{0}{\includegraphics{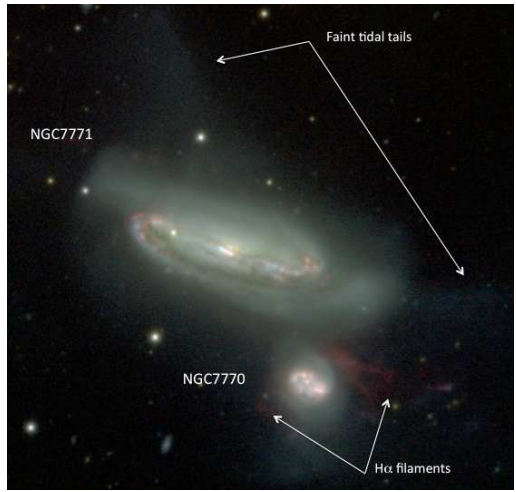}}}

     \caption[]{ALFOSC false-color RGB image of the NGC~7771+NGC~7700
       system constructed
       using the SDSS broad-band $g$ and $r$  images in blue and green,
       respectively, and the
       narrow-band H$\alpha$+[N\,{\sc ii}] image in
       red. Orientation is north up, east to the left. The
       approximate FoV is $210\arcsec \times 200\arcsec$. 
We also mark some morphological
       features discussed in  Section~\ref{sec:gas}.}
        \label{fig:NOT_RGB}
    \end{figure}

In this work we use optical imaging and integral field spectroscopy (IFS)
to study the impact of the minor merger on the star formation properties of the 
individual galaxies of the NGC~7770+NGC~7771 system. We use these
observations to test the predictions of  minor merger simulations. The primary
galaxy NGC~7771 is an
SB(s)a galaxy, the satellite galaxy NGC~7770 is classified as
an S0/a (see Fig.~\ref{fig:NOT_RGB}), and they have an approximate
stellar mass ratio of 10:1 \citep{PereiraSantaella2011}. They are in a group 
with NGC~7769 to the northwest of the system and a small
galaxy to the west of NGC~7771, with all of
them being embedded in a common neutral hydrogen envelope
\citep{Nordgren1997}.  Throughout this work we assume a common
distance of 60\,Mpc.

\section[]{Observations}\label{sec:observations}

We used  the Potsdam Multi Aperture Spectrograph 
\citep[PMAS, ][]{Roth2005}  in the PMAS fiber Package mode
\citep[PPAK, ][]{Kelz2006} on the 3.5m telescope in Calar Alto to
observe the NGC~7771+NGC~7770 system. This was part of the PPAK IFS
Nearby Galaxies Survey (PINGS).   
Very briefly, we used the V300 grating
to cover the $3700-7100$\AA \, spectral range with a spectral resolution
of $10$\AA.  We took three pointings with a  dithered pattern, 
which allowed us to resample the 
PPAK  2.7\arcsec-diameter fiber to a final mosaic with  a 2\arcsec \,
spaxel and a field of view (FoV) of $148\arcsec \times 130\arcsec$. 
We reduced the data following  \cite{Sanchez2006}.
Before constructing the  maps of the emission lines  we fitted the stellar
continuum with the {\sc fit3d} routine  \citep{Sanchez2007}.  
We used the
\cite{Bruzual2003} models to generate single 
stellar populations (SSP) with a Salpeter IMF covering a range of
ages
and metallicities.
We refer the reader to
\citet{Sanchez2011} for more details.  We finally
 subtracted the fitted stellar continuum and  used 
the observed H$\alpha$/H$\beta$ line ratio to correct for extinction
the line maps  on a spaxel-by-spaxel basis \citep[see][for a full description]{RosalesOrtega2010}. 


We  also obtained imaging with the Andalucia Faint Object
Spectrograph and Camera (ALFOSC) on the 2.5m Nordic Optical  Telescope
(NOT) in El Roque de los Muchachos. 
We used the SDSS broad-band 
$g$ and $r$  filters and an 
H$\alpha$+[N\,{\sc ii}] narrow-band filter ($\lambda_{\rm c}=
6653$\AA \, and $\Delta \lambda=55$\AA \, at full width half maximum FWHM). 
The instrument has a
0.19\arcsec/pixel plate scale and the observations were obtained under
photometric and  extremely good seeing  ($\sim 0.6\arcsec$,
FWHM as measured
from stars in the FoV)  conditions. We
followed standard  techniques to reduce and calibrate the data,
as well as to subtract the continuum from the H$\alpha$+[N\,{\sc ii}]
image and compute the H$\alpha$ emission. 

\section[]{Results}\label{sec:results}

\subsection{Optical continuum and ionized gas properties}\label{sec:gas}
The deep ALFOSC optical continuum images allow us to get a detailed
view of the morphology of the system, especially of the lesser known
galaxy NGC~7770. Although NGC~7770 has been classified  as
an S0/a,  it shows a faint 
spiral structure with bright knots located in a ring of star formation
(Fig.~\ref{fig:NOT_RGB}). 
There are also other interesting features seen in the optical
continuum emission such as a truncated stellar disk in NGC~7771 and faint
short tidal tails in both galaxies. Such continuum
tail features are predicted by simulations of minor mergers right
after the first 
passage \citep[][]{Cox2008}.

\begin{figure*}
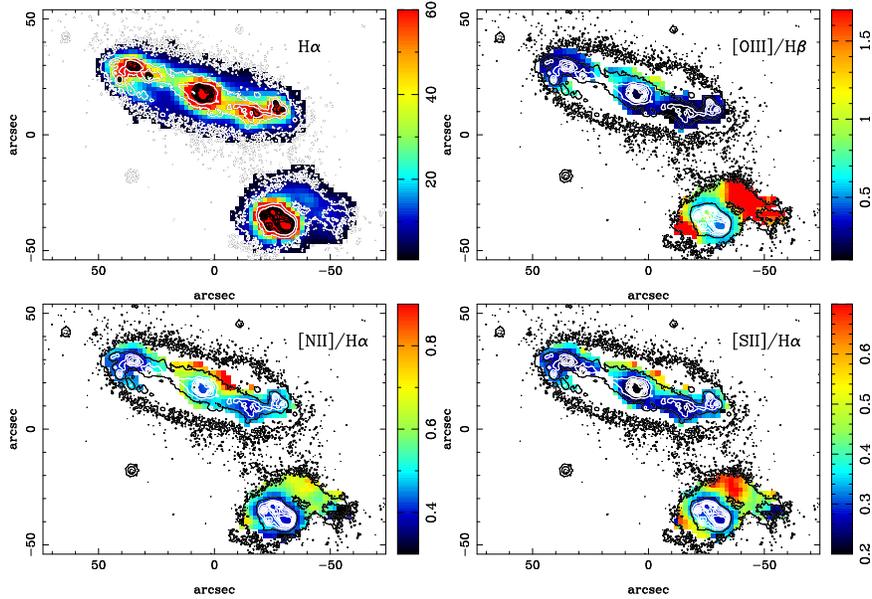


\hspace{0.5cm}
\resizebox{0.32\hsize}{!}{\rotatebox[]{0}{\includegraphics{figure2a.ps}}}
\resizebox{0.32\hsize}{!}{\rotatebox[]{0}{\includegraphics{figure2b.ps}}}

\hspace{0.5cm}
\resizebox{0.32\hsize}{!}{\rotatebox[]{0}{\includegraphics{figure2c.ps}}}
\resizebox{0.32\hsize}{!}{\rotatebox[]{0}{\includegraphics{figure2d.ps}}}

     \caption[]{PPAK spectral maps  (color images) of the 
observed  H$\alpha$ emission (units of $10^{-16}\,{\rm
    erg\,cm}^{-2}\,{\rm s}^{-1}\,{\rm arcsec}^{-2}$) and  
the extinction-corrected [O\,{\sc iii}]$\lambda$5007/H$\beta$, 
[N\,{\sc ii}]$\lambda$6583/H$\alpha$, and [S\,{\sc
  ii}]$\lambda\lambda$6717,6731/H$\alpha$ line ratios. The 
contours  (in a square root scale) are the ALFOSC continuum-subtracted 
H$\alpha$ emission.}
        \label{fig:PPAK_maps}
    \end{figure*}

\begin{figure*}
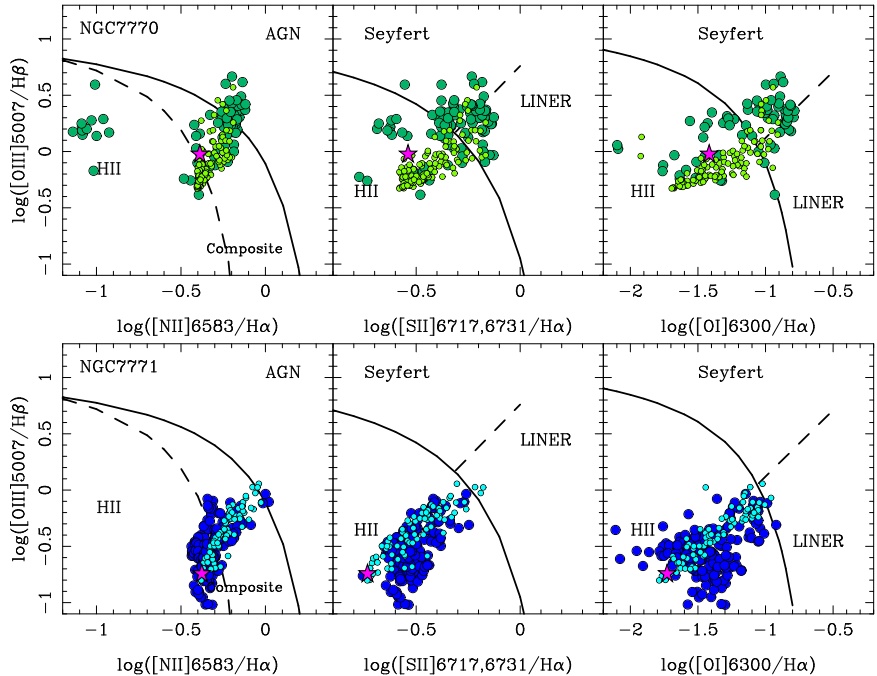


\resizebox{0.65\hsize}{!}{\rotatebox[]{-90}{\includegraphics{figure3a.ps}}}

\vspace{-3.5cm}
\resizebox{0.65\hsize}{!}{\rotatebox[]{-90}{\includegraphics{figure3b.ps}}}

\vspace{-3.7cm}
     \caption[]{Spatially-resolved diagnostic diagrams. Each
       measurement corresponds to one spaxel. The star-like
symbols are the nuclear line ratios, and the small and large dots
are spaxels at projected 
galactocentric distances of $R<12\arcsec$ and $R>12\arcsec$,
       respectively. 
The solid curves
are ``maximum starburst lines'' \citep{Kewley2001}
and the dashed curves the empirical separation between Composites
and H\,{\sc ii} regions and Seyferts and LINERs \citep{Kauffmann2003BPT,
  Kewley2006}. The spaxels of NGC~7770 (upper panel) 
in the LINER/Seyfert and AGN
regions can also be produced by shocks \citep[see text and
also][]{Allen2008}. }\label{fig:PPAK_BPTs}  
    \end{figure*}

Both galaxies in the system
show extended H$\alpha$ emission over several kpc (Fig.~\ref{fig:PPAK_maps}). 
In NGC~7771 the H$\alpha$ emission arises from the nuclear region of the galaxy,
which is in fact a ring of star 
formation \citep[see][and references therein]{AAH10}, from
bright H\,{\sc ii} regions at the end of the bar and in
the spiral arms, and from the disk of the galaxy. NGC~7770 also shows 
a circumnuclear  ring of star formation
with an approximate diameter 
of 20\arcsec \, ($\sim 5.7\,$kpc). 
The rings of star formation in both galaxies show 
optical line ratios typical of H\,{\sc ii} regions, as can be seen 
from the PPAK spectral maps (Fig.~\ref{fig:PPAK_maps}) and the
spatially-resolved diagnostic diagrams (Fig.~\ref{fig:PPAK_BPTs}). 
For the spaxels in both galaxies showing H\,{\sc ii} region-like 
line ratios, the differences in the observed [O\,{\sc iii}]/H$\beta$
line ratios are readily explained by differences in the gas-phase oxygen
abundances (see Section~\ref{sec:abundances}).
 
The most striking
feature are the filaments of diffuse  H$\alpha$ emission in the
interface region between the two galaxies to the northwest of
NGC~7770, and also to the south of the galaxy (see Fig.~\ref{fig:NOT_RGB}). 
These H$\alpha$
filaments are not detected in the optical continuum emission 
(Fig.~\ref{fig:NOT_RGB}) indicating
that they are mostly made of gas. The northwest filaments extend out
to projected  galactocentric distances of approximately
6\,kpc. Minor-merger simulations  \citep[][]{Cox2008} predict that  
gas is likely to have been
cleared out from the primary galaxy extended disk after  the first
close passage of the satellite galaxy. 
The H$\alpha$ filaments of NGC~7770 and a few spaxels tracing the
diffuse H$\alpha$ emission around the ring of star formation of
NGC~7771 \citep[see also][]{AAH10} 
display  elevated optical line ratios. This is better
seen in the spatially-resolved diagnostic diagrams
(Fig.~\ref{fig:PPAK_BPTs}) where 
a large fraction of the outer (i.e., projected galactocentric distances
$>12\arcsec \sim 3.5\,$kpc) H$\alpha$ emitting regions of  NGC~7770
are  in the composite/AGN region or the LINER/Seyfert region. 
The observed line ratios for the filament spaxels,
even those falling in the LINER/Seyfert region, can be reproduced with the
shock+precursor models of \citet{Allen2008} with shock velocities of
$300-400\,{\rm 
  km\,s}^{-1}$ (their figures 31, 32, and
33). This could be understood if gas, from either NGC~7771 or 
NGC~7770, is being stripped and excited by shocks induced by tidal 
forces. In NGC~7771 there is
no variation of the  optical line ratios with the galactocentric
distance.

\subsection{Luminosity-weighted stellar ages and metallicities}\label{sec:ages}
The luminosity-weighted ages and metallicities of 
stellar populations can give us
hints about the {\it recent}  star formation history of
galaxies. To ensure an appropriate modelling of the stellar
populations we used a Voronoi binning method 
\citep{Cappellari2003} and the {\sc pingsoft}
\citep{RosalesOrtega2011} software to define regions (also known as voxels) 
in the PPAK data cube with extracted spectra of  
sufficient signal-to-noise (S/N) ratios. We targeted S/N ratios of 25
over a 
 spectral region centered at
5200\,\AA \, with a 100\,\AA-width. We then used the {\sc fit3d}
routine to fit SSPs to the entire spectrum of each voxel
with a $\chi^2$ minimization
technique. We generated a grid of templates using 
the \citet{Bruzual2003} models with a range of ages (5\,Myr, 25\,Myr, 
100\,Myr, 640\,Myr, 1.4\,Gyr, 2.5\,Gyr, 5\,Gyr, 13 Gyr, and 17 Gyr) 
and 3 metallicities ($Z=0.008$, 0.02, and
0.05), and allowed for the templates to be attenuated. 
In general it was not possible to distinguish among  
the few best-fitting models. We therefore 
estimated the age and metallicity of each extracted spectrum
by averaging the ages and
metallicities of these  models, weighted by their 
corresponding $\chi^2$ values.
For the following discussion we grouped the stellar
populations in young and post-starburst with ages of $<$1\,Gyr,
intermediate with ages of $1-5\,$Gyr, and old with ages $>$5\,Gyr.

\begin{figure}
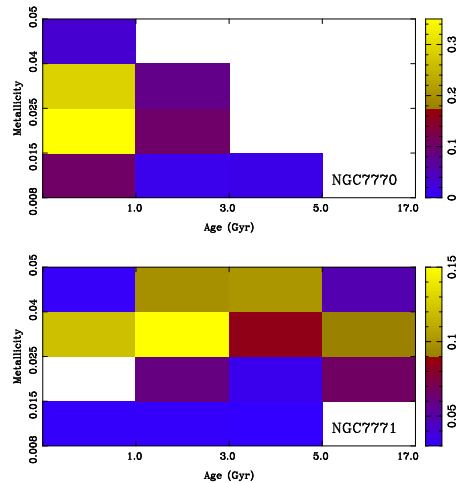

\hspace{0.95cm}
\resizebox{0.7\hsize}{!}{\rotatebox[]{-90}{\includegraphics{figure4a.ps}}}

\vspace{-1.2cm}
\hspace{0.95cm}
\resizebox{0.7\hsize}{!}{\rotatebox[]{-90}{\includegraphics{figure4b.ps}}}

\vspace{-1.3cm}
     \caption[]{Distributions of the 
       ages and metallicities of the spatially-resolved
       (i.e., corresponding to the extracted spectra for the defined
       voxels, see text) modelled SSPs. The colors represent the 
fractional contribution of the SSP bins
       (luminosity-weighted) to the  observed optical 
       emission ($3700-7100\,$\AA) of each galaxy. }
        \label{fig:PPAKNOT_stellar}
    \end{figure}

The optical light of the primary galaxy is mostly
dominated by intermediate and  old stellar populations (see
Fig.~\ref{fig:PPAKNOT_stellar}) with super-solar metallicities ($Z>0.025$). 
There is a small contribution ($\sim 20\%$) 
from stellar populations with ages of $<$1\,Gyr, which 
are mostly located in the circumnuclear 
ring of star formation. This agrees
with  the presence of  post-starburst
stellar populations in the nuclear region of NGC~7771 found by
previous works
\citep[see][]{Davies1997, AAH10}. 
In the satellite galaxy NGC~7770 the young stellar populations 
($<1$Gyr) with solar and super-solar metallicities ($Z=0.015-0.04$)
contribute as much as $\sim 80$\% of the optical
light. 

The presence of a post-starburst stellar
  population has often been interpreted as the result of a past
  interaction \cite[e.g.,][]{Liu1995}. The fact that the optical
  emission of  NGC~7770 is dominated by this population suggests that 
the galaxies have already experienced first passage. This likely
triggered a strong burst of star formation in  
the satellite galaxy less than 1\,Gyr ago, as predicted by the simulations of
\citet{Cox2008}. The origin of the 
post-starburst population 
in NGC~7771
 is less clear since these minor-merger models do
not predict a 
strong star formation enhancement in the primary.  
Based on the neutral hydrogen morphology,  
\citet{Nordgren1997} suggested that 
NGC~7771 and NGC~7769,  which have a 2:1 mass ratio, appear to be
having a prograde--retrograde  
interaction, with NGC7769 being the retrograde one.  Such
an interaction, while not as efficient at triggering strong episodes of star
formation, may explain the circumnuclear burst of star formation in NGC~7771.

\subsection{Current Star Formation Activity}\label{sec:sfr}
Next we  explore the  impact of the  interaction on the
current SFR of the galaxies. For a 10:1 stellar
mass ratio \citet[][]{Cox2008} predict
only a mild total enhancement ($< 2 \times$)  in the global SFR of the
system when compared to the sum of  SFRs of the individual galaxies if
they were isolated. In fact, most of the SFR 
enhancement takes place in the satellite galaxy because it is that experiencing
significant tidal forces. We can test this prediction for the
individual galaxies of this system since in the local Universe star-forming
galaxies define a sequence of increasing SFR for increasing stellar masses
\citep[e.g.,][and references therein]{Salim2007}.

The integrated extinctions $A_V$ of the galaxies ($1.1\,$mag and 
$2.5\,$mag for NGC~7770 and NGC~7771, respectively, from the
 H$\alpha$/H$\beta$ ratio) 
indicate that we can use the SFR recipe of  
\citet{Kennicutt2009}. It combines the observed (not corrected
for $A_V$) H$\alpha$ luminosity and the IR
 luminosity to account for  the unobscured and
the obscured SFRs, respectively. 
The observed H$\alpha$ luminosities are  
$L({\rm H}\alpha) = 2.7 \times
10^{41}\,{\rm erg \,s}^{-1}$ and  $4.2 \times
10^{41}\,{\rm erg \,s}^{-1}$ for NGC~7770 and NGC~7771, respectively,
and the  
IR luminosities  are $\log (L_{\rm IR}/{\rm
  L}_\odot) = 10.8$ and $11.3$ \citep[][]{PereiraSantaella2011}. 

There is some evidence of the presence of an AGN in NGC~7770 
both from the composite nuclear (i.e.,
star-forming/AGN) [N\,{\sc   ii}]/H$\alpha$  line ratio
(Fig.~\ref{fig:PPAK_BPTs}) and  the hot dust emission detected in the
Spitzer spectrum   \citep{AAH12}.  To compute the SFR we then 
need to subtract the AGN  contribution from the IR luminosity.
    For a Kroupa IMF we obtained  SFR(total)$= 4.5\,{\rm
M}_\odot \,{\rm yr}^{-1}$ and $12.8\,{\rm
M}_\odot \,{\rm yr}^{-1}$ for NGC~7770 and NGC~7771,
respectively. These  are similar, within the uncertainties of
the SFR recipes ($\pm0.3\,$dex or better),  
to those obtained from the extinction-corrected
L(H$\alpha$). We note that the SFR recipe used here assumes that
approximately half of the dust heating arises from stars older than
30\,Myr. The IR-based calibration of \citet{Kennicutt1998},
which assumes that the dust is only heated by these young stars,
results in SFRs approximately two times higher.

The specific SFR (SSFR$={\rm SFR}/M_*$) using 
stellar masses computed with a near-IR mass-to-light ratio
\citep[][]{Bell2001, PereiraSantaella2011} are $\log ({\rm
  SSFR/yr}^{-1})=-9.6$ and $-10.4$ for NGC~7770 and NGC~7771,
respectively. If there was no AGN in NGC~7770 then the calculated
  SFR and SSFR would be  lower limits. We find that NGC~7771 falls   
on the high mass end of  the local star-forming sequence. 
Composite galaxies and AGN 
 follow a similar star-forming sequence  but
extending to higher stellar masses \citep[][their figure~18]{Salim2007}. 
The satellite galaxy NGC~7770 shows an elevated
SSFR and lies just off of the
distinct  region occupied by local composite galaxies. 
This suggests that not only does NGC~7770
show evidence of a post-starburst stellar population, but that it is
also currently undergoing a
galaxy-wide burst of star formation. 
These results agree with  predictions of minor merger simulations 
as well as with 
statistical observational studies of minor mergers 
\citep{Woods2007,Ellison2008}.

\subsection{Gas-phase oxygen abundances}\label{sec:abundances}
Interacting galaxies are believed to contribute to the observed
scatter in the local mass-metallicity
relation \citep{Tremonti2004}. Specifically, 
local luminous and ultraluminous IR galaxies,
which include a large fraction of gas-rich mergers, 
show lower nuclear  abundances than local
emission-line galaxies of similar luminosity and mass \citep{Rupke2008}.
Simulations of interacting galaxies
show that it is possible to decrease the nuclear metallicity
although the gas  content of the galaxies  plays
a role \citep{Torrey2012}. 

\begin{figure}

\hspace{0.3cm}
\resizebox{0.9\hsize}{!}{\rotatebox[]{-90}{\includegraphics{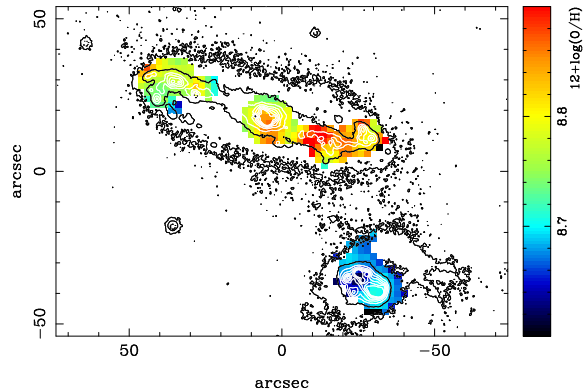}}}
\vspace{-1.5cm}
     \caption[]{Map of oxygen abundances derived with the O3N2 index
     and the calibration of \citet{Pettini2004} only for spaxels
   with no evidence of shock excitation ([O\,{\sc i}]$\lambda$6300/H$\alpha
   <0.06$). Contours are as in Fig.~\ref{fig:PPAK_maps}.}
        \label{fig:metallicities}
    \end{figure}

We used the calibration of  \citet{Pettini2004} based on the
O3N2 index to derive the oxygen abundances of
NGC~7770 and NGC~7771 on a spaxel-by-spaxel basis,  but only
for regions with [O\,{\sc i}]/H$\alpha$$<$0.06. In doing so  we
ensured that most of 
the line emission used to compute the abundances is produced by
photoionization in H\,{\sc ii} regions rather than by other mechanisms. 
The oxygen abundances
of the regions of NGC~7770 are lower that those of NGC~7771
(Fig.~\ref{fig:metallicities}) as 
expected from the different stellar masses and in agreement with the
SSP modelling. The 
distribution of the oxygen abundance in NGC~7771 is rather complex
with the H\,{\sc ii} regions at the east end of the bar having
apparently lower abundances than those to the west.
The abundance in the central regions of NGC~7770 
appears to be about 0.1dex lower than in the bright H\,{\sc ii} region
to the southwest of the nucleus and in the outskirts of the galaxy. This
may occur if gas from the outskirts of the
satellite galaxy has been channeled to the nuclear region as the result of the
interaction. However the suspected composite nature of the
nucleus of NGC~7770  may complicate this interpretation.

\section{Conclusions}\label{sec:conclusions}

We presented PPAK spatially-resolved optical emission line
ratio and abundance maps and diagnostic diagrams as
well as ALFOSC deep optical imaging of the
NGC7771+NGC7770 system to study the impact of the 
interaction on the star formation properties of the individual
galaxies. The galaxies
have a $\sim$ 10:1 stellar mass ratio and thus are classified as a minor
merger. Numerical simulations of minor mergers 
 predict little global
enhancement of the star formation properties and mostly affecting the
satellite galaxy. 
We find that the satellite galaxy in the system NGC~7770
is indeed experiencing a galaxy-wide starburst with most of 
of its optical emission  being produced by young and 
post-starburst stellar populations
($<1\,$Gyr). This galaxy shows an enhanced SSFR with respect
to the distinct region occupied by composite objects in the 
 local star-forming sequence. 
We also detected H$\alpha$ emitting gas filaments in the outskirts of 
NGC~7770 at projected 
galactocentric distances greater than $\sim 4\,$kpc and out to 6\,kpc. 
These filaments appear to be made of gas that has been stripped from one of
the two  galaxies and is being excited by shocks.
Overall, our findings support the picture that the minor merger induced
one or several episodes of star formation  in the satellite galaxy
NGC~7770 after the first
close passage. Such  interaction-induced effects are not observed in
the primary galaxy of the system.

We thank a referee for comments that helped improve the paper
and  M. Cappellari 
for interesting discussions. 
A.A.-H. acknowledges  finantial support from the Spanish Plan Nacional
grant AYA2010-21161-C02-01 and the Universidad de Cantabria
 AGL program,  A.I.D. from 
grant AYA2010-21887-C04-03, and  F.F.R.-O. from 
the Mexican National Council for Science and Technology (CONACYT) 
through the programme Estancias Posdoctorales y
Sab\'aticas al Extranjero para la Consolidaci\'on de Grupos de
Investigaci\'on, 2010-2011. M.P.-S. is funded by an ASI fellowship
under contract  I/005/11/0. 

\label{lastpage}

\end{document}